\documentclass[11pt]{article}

\usepackage{amsmath,amsfonts,amsthm}
\usepackage[breaklinks,pdfpagemode=None,pdfview=FitH,pdfstartview=FitH,citebordercolor={0 0 1},linkbordercolor={0 0 1},urlbordercolor={0 0 1},pagebordercolor={0 0 1}]{hyperref}

\input colordvi    %% contains list of colours and macro explanation
\newcommand{\m}{\mu}
\begin{document}

\title{THE EXACT RENORMALIZATION GROUP}

\author{
     P. K. Mitter\thanks {e-mail: pkmitter@LPTA.univ-montp2.fr} \\
     Laboratoire de Physique Th\'eorique et Astroparticules
     \thanks{Laboratoire Associ\'e au CNRS, UMR 5207}\\
     Universit\'e Montpellier 2\\
     Place E. Bataillon, Case 070\\ 
     34095 Montpellier Cedex 05 France
}    

\date{ Contribution to the {\it Ecyclopedia of Mathematical Physics}, J-P.
Francoise, G.Naber and T.S. Tsun, eds. (Elsevier,2006.)\\
\today
}

\maketitle

\begin{abstract}
This is a very brief introduction to Wilson's Renormalization
Group with emphasis on mathematical developments.
\end{abstract}

\section{Introduction} \label{sec-introduction}
\setcounter{equation}{0}

The renormalization group (RG) in its modern form was invented by K.G.
Wilson in the context of statistical mechanics and euclidean quantum field 
theory (EQFT). It offers the deepest understanding of renormalization in 
quantum field theory by connecting EQFT with the the theory of second order 
phase transition and associated critical phenomena. Thermodynamic functions
of many statistical mechanical models ( the prototype being the Ising model
in two or more dimensions) exhibit power like singularities as the temperature
approaches a critical value. 
One of the major triumphs of the Wilson RG was the prediction of 
the exponents ( known as critical exponents) associated to these singularities.
Wilson's fundamental contribution was to realize that many length scales
begin to cooperate as one approaches criticality and that one should 
disentangle them and treat them one at a time. This leads to an iterative 
procedure known as the renormalization group. Singularities and critical 
exponents then arise from a limiting process. Ultraviolet singularities 
of field theory can also
be understood in the same way. Wilson reviews this in [WK 1974] and gives the 
historical genesis of his ideas in [W 1983].

The early work in the subject was heuristic in the sense that clever but
uncontrolled approximations were made to the exact equations often with 
much success. Subsequently authors with mathematical bent began to use the
underlying ideas to prove theorems: Bleher
and Sinai [BS 1975] in work related to critical phenomena and
Benfatto, Cassandro, Gallavotti et al 
[BCGNOPS 1978, 1980] in the construction of super-renormalizable quantum
field theories. The subject saw further mathematical development
in the work of Gawedzki and Kupianen [ GK 1980, 1983, 1984, 1985]  and Balaban 
[1982a, 1982b]. Balaban in a series  papers ending in [B 1989] proved
a basic result on the continuum limit of Wilson's lattice gauge theory. Brydges
and Yau [BY 1990] simplified the mathematical treatment of the renormalization
group  for a class of models and this has lead to further systemization and 
simplification in the work of Brydges, Dimock and Hurd [BDH 1998] and 
Brydges, Mitter and Scoppola [BMS  2003]. Another method which has been
intensely developed during the same historical period is based on phase cell 
expansions: Feldman, Magnen, Rivasseau and S\'en\'eor [FMRS 1986, 1987] 
developed the early phase cell ideas of Glimm and Jaffe and were able to
prove independently many of the results cited earlier. Although these 
methods share many features of the Wilson RG they are different in 
methodology and thus remain outside of the purview of the present exposition.  

A somewhat different line of 
development has been the use of the RG to give simple proofs of perturbative
renormalizability of various quantum field theories: Gallavotti and Nicolo
[GN 1985] via iterative methods and Polchinski [P 1984] who exploited a 
continuous version of the RG for which Wilson [W 1974] had derived a 
non-linear differential equation. Polchinski's work has been extended to 
non-abelian gauge theories by Kopper and Muller [KM 2000].

Finally it should be mentioned that apart from quantum field theory and
statistical mechanics the RG method has proved fruitful in other domains.
An example is the study of interacting fermionic systems in condensed 
matter physics ( see the articles of Mastropietro and Salmhofer in 
this volume). In the rest of this article our focus will be on quantum field 
theory and statistical mechanics.

\section{The RG as a Discrete Semigroup}\label{sec-ex}
\setcounter{equation}{0}

We will first define a discrete version of the RG and consider its continuous 
version later. As we will see the RG is really a {\it semigroup}. Thus 
calling it a group is a misnomer.
 
Let $\phi$ be a Gaussian random field (see e.g. Gelfand and Vilenkin, [GV 1964]
for random fields) in ${\bf R}^{d}$. Associated to it there is a positive 
definite function which is identified as its covariance.
In quantum field theory one is interested in the covariance  
\begin{equation}
E(\phi(x)\phi(y)) =const.|x-y|^{-2[\phi]}=
\int_{{\bf R}^d} dp\> e^{ip.(x-y)}\> \frac {1}{|p|^{d-2[\phi]}}  
\label{eq-ex.1}
\end{equation}
Here $[\phi]>0$ is the (canonical) {\it dimension} of the field, which for the
standard massless free field is $[\phi]=\frac{d-2}{2}$. The latter is positive
for $d > 2$. However other choices are possible but in EQFT they are 
restricted by Osterwalder-Schrader positivity.
This is assured if $[\phi]= \frac{d-\alpha}{2}$ with $0<\alpha \le 2$. If
$\alpha < 2$ we get a generalized free field.
 
Observe that the covariance is singular
for $x=y$ and this singularity is responsable for the ultraviolet 
divergences of quantum
field theory. This singularity has to be initially cutoff and  
there are many ways to do this. A simple way to do this is as follows. 
Let $u(x)$ be a smooth, rotationally invariant, 
{\it positive definite} function of fast decrease. Examples of such functions
 are legion. Observe that
 
\begin{equation}  \label{eq-ex.1a}
|x-y|^{-2\phi}
=const.\int_{0}^{\infty}\frac{dl}{l}\> l^{-2[\phi]}\> u(\frac{x-y} {l})
\end{equation}
as can be seen by scaling in $l$. We define the {\it unit} ultraviolet 
cutoff covariance
$C$ by cutting off at the lower end point of the $l$ integration  
( responsable for the singularity at $x=y$) at $l=1$, 
 
\begin{equation}
C(x-y)=\int_{1}^{\infty}\frac{dl}{l} \> l^{-2[\phi]}\> u(\frac{x-y}{l})
\label{eq-ex.2}
\end{equation}
$C(x-y)$ is positive definite and everywhere smooth. Being positive definite
it qualifies as the covariance of a Gaussian probability 
measure denoted $\mu_{C}$ on a function space $\Omega$ ( which it is not 
necessary to specify  any further). The covariance 
$C$ being smooth implies  that the sample fields of the measure are 
$\mu_{C}$ almost everywhere  sufficiently differentiable. 

\vskip0.3cm

\noindent {\it Remark} : Note that more 
generally we could have cutoff the lower end point singularity in 
(\ref{eq-ex.1}) at any $\epsilon > 0$. The $\epsilon$-cutoff covariance is 
related to the unit cutoff covariance by a scale transformation 
( defined below) and we will exploit this relation later. 

\vskip0.3cm

Let $L > 1 $ be any real number. We define a scale transformation 
$S_{L}$ on   fields $\phi$ by

\begin{equation}
S_{L}\phi(x)=L^{-[\phi]}\> \phi(\frac{x}{ L}) \label{eq-ex.3}
\end{equation}
on covariances by

\begin{equation}
S_{L}C(x-y)=L^{-2[\phi]}\> C(\frac{x-y}{ L})
\end{equation}
and on functions of fields $F(\phi)$ by

\begin{equation}
S_{L}F(\phi)=F(S_{L}\phi)
\end{equation}
The scale transformations form a {\it multiplicative group} : 
$S_{L}^{n}=S_{L^{n}} $.

Now define a
{\it fluctuation covariance } $\Gamma_{L}$ 

\begin{equation}
\Gamma_{L} (x-y)=\int_{1}^{L}\frac{dl}{l} \> l^{-2[\phi]}\> u(\frac{x-y}{l})
\label{eq-ex.4}
\end{equation}
$\Gamma_{L} (x-y)$ is smooth, positive definite and is of fast decrease on
scale $L$. It generates a key scaling decomposition 

\begin{equation}
C(x-y)=\Gamma_{L} (x-y)\> + S_{L} C(x-y)
\label{eq-ex.5}
\end{equation}
Iterating this we get 

\begin{equation}
C(x-y)=\sum_{n=0}^{\infty}\> \Gamma_{n}(x-y)  
\label{eq-ex.6}
\end{equation}
where
\begin{equation}\label{eq-ex.7}
\Gamma_{n}(x-y)=S_{L^{n}}\Gamma_{L} (x-y)=
L^{-2n[\phi]}\> \Gamma_{L} (\frac{x-y}{L^{n}}) 
\end{equation}
The  $\Gamma_{n}(x-y)$ are of fast decrease on scale $L^{n+1}$.

Thus (\ref{eq-ex.6}) achieves the decomposition into a sum over increasing
length scales as desired. Being positive definite the $\Gamma_{n}$ 
qualify as covariances of Gaussian probability measures , and therefore
$\mu_{C}=\> \bigotimes\limits_{n=0}^{\infty}\> 
\mu_{\Gamma_{n}}$. 
Correspondingly introduce
a family of independent Gaussian random fields $\zeta_{n}$, called
{\it fluctuation fields}, distributed  according to $\mu_{\Gamma_{n}}$ .
Then

\begin{equation}
\phi=\> \sum_{n=0}^{\infty}\> \zeta_{n} \label{eq-ex.9} 
\end{equation}
Note that the fluctuation fields $\zeta_{n}$ are slowly varying over  length
scales $L^{n}$. In fact an easy estimate using
a Tchebycheff inequality shows that for any $\gamma >0$ 

\begin{equation}
|x-y|\le L^{n} \Rightarrow 
\mu_{C}\Big(|\zeta_{n}(x) - \zeta_{n}(y)|\ge \gamma \Big)
\le const. \gamma^{-2}
\end{equation}
which reveals the slowly varying nature of $\zeta_{n}$ on scale $L^{n}$.
(\ref{eq-ex.9}) is an example of a multiscale decomposition of a 
Gaussian random field.

The above implies that the $\mu_{C}$ integral of a 
function can be written as 
a multiple integral over the fields $\zeta_{n}$ . We calculate it
by  integrating out the fluctuation fields $\zeta_{n}$ step by step 
going from shorter to longer length scales. This can be accomplished by the 
iteration
of a single transformation $T_{L}$, a {\it renormalization group 
transformation}, as follows. Let $F(\phi)$ be a function of fields. Then
we define a RG transformation  $F\rightarrow T_{L} F$ by

\begin{equation}\label{eq-ex.10}
(T_{L} F)(\phi)= S_{L} \mu_{\Gamma_{L}}*F(\phi)= 
\int d\mu_{\Gamma_{L}}(\zeta)\> F(\zeta + S_{L}\phi) 
\end{equation}
 {\it Thus the renormalization group transformation consists of a 
convolution with 
the fluctuation measure followed by a rescaling.}\\
\\ {\it Semigroup property} : The discrete RG transformations form a 
{\it semigroup}
\begin{equation}\label{eq-ex.11}
T_{L}T_{L^{n}}  = T_{L^{n+1}}
\end{equation}
for all $n\ge 0$.\\
\\To prove  this we must first see how scaling commutes with convolution with
a measure. We have the property :

\begin{equation}
\mu_{\Gamma_{L}}* S_{L}F= S_{L}\mu_{S_{L}\Gamma_{L}}*F \label{eq-ex.12}
\end{equation}
To see this observe first that if $\zeta$ is a Gaussian random field 
distributed with covariance $\Gamma_{L}$ then the Gaussian field 
$S_{L}\zeta$ is distributed according to $S_{L}\Gamma_{L}$. This can be checked
by computing the covariance of $S_{L}\zeta$. Now the lefthand side of
(\ref{eq-ex.12}) is just the integral of $F(S_{L}\zeta + S_{L}\phi)$ with 
respect to $ d\mu_{\Gamma_{L}}(\zeta)$. By the previous observation this is
the integral of $F(\zeta + S_{L}\phi)$ with respect to 
$ d\mu_{S_{L}\Gamma_{L}}(\zeta)$, and the latter is the right hand side of 
(\ref{eq-ex.12}).\\
\\Now we can check the semigroup property trivially :

\begin{align}\label{eq-ex.13} 
T_{L}T _{{L}^{n}}F =S_{L} \mu_{\Gamma_{L}}* S_{L^{n}} \mu_{\Gamma_{L^{n}}}* F
 & = S_{L}S_{{L}^{n}}\mu_{S_{L^{n}}\Gamma_{L}}*\mu_{\Gamma_{L^{n}}}* F \cr 
=S_{L^{n+1}}\mu_{\Gamma_{L^{n}}+ S_{L^{n}}\Gamma_{L}}*F 
& =S_{L^{n+1}}\mu_{\Gamma_{L^{n+1}}}*F  \cr
=T_{L^{n+1}}F 
\end{align}
We have used the fact that $\Gamma_{L^{n}}+ S_{L^{n}}\Gamma_{L}=
\Gamma_{L^{n+1}}$. This
is because $S_{L^{n}}\Gamma_{L}$ has the representation (\ref{eq-ex.4}) with 
integration interval changed to $[L^{n}, L^{n+1}]$.\\
\\ We note some properties of $T_{L}$. $T_{L}$ has an unique invariant 
measure, namely $\mu_{C}$ : For any bounded function $F$

\begin{equation}
\int d\mu_{C} T_{L} F = \int d\mu_{C} F  
\label{eq-ex.14}
\end{equation}
To understand (\ref{eq-ex.14}) recall the earlier observation 
that if $\phi$ is distributed according to the covariance $C$, then
$ S_{L}\phi $ is distributed according to $S_{L}C $. By (\ref{eq-ex.5})
$ \Gamma_{L} +S_{L}C = C $. Therefore

\begin{align}\label{eq-ex.15}
\int d\mu_{C} T_{L} F & = \int d\mu_{C} 
S_{L}\mu_{\Gamma_{L}}*F \cr
& = \int d\mu_{S_{L}C}\mu_{\Gamma_{L}}*F \cr
& = \int d\mu_{C} F
\end{align}
The uniqueness of the invariant measure follows from the fact that the
semigroup $T_{L}$ is realized by a convolution with a probability measure
and therefore is positivity improving: 
$F\ge 0,\> \mu_{C}\> a.e. \Rightarrow T_{L} F > 0, \> \mu_{C} \> a.e.$

Finally note that $T_{L}$ is a {\it contraction semigroup } on 
$L^{p}(d\mu_{C})$ for  $1\le p < \infty$. To see this note that since
$ T_{L}$ is a convolution with a probability measure :
$T_{L}F= \mu_{S_{L^{-1}}\Gamma_{L}}*S_{L} F$ we have via H\"older's 
inequality
$|T_{L}F|^{p} \le T_{L}|F|^{p}$. Then use the fact that 
$\mu_{C}$ is an invariant measure.\\
\\{\it Eigen functions} : Let $:p_{n,m}:(\phi (x))$ be a $ C$
Wick ordered local monomial of $m$ fields with $n$ derivatives.
Define $P_{n,m}(X)= \int_{X} dx \> :p_{n,m}:_{C}(x)$.
The $P_{n,m}(X)$ play the role of
eigenfunctions of the RG transformation $T_{L}$ {\it upto a scaling
of volume}:

\begin{equation}
T_{L} P_{n,m}(X) = L^{d-m[\phi]-n} P_{n,m} (L^{-1} X) \label{eq-ex.16}
\end{equation}
Because of the scaling in volume the $P_{n,m}(X)$ are not true eigenfunctions.
Nevertheless they are very useful because they play an important role in the 
analysis of the evolution of the dynamical system which we will later
associate with $T_{L}$.
They are classified as expanding ({\it relevant}), contracting 
({\it irrelevant}) or central ({\it marginal}) depending on whether the 
exponent of $L$ on the
right hand side of \ref{eq-ex.16} is positive, negative or zero. This
depends of course on the space dimension $d$ and the 
field dimension $[\phi]$.\\
\\Gaussian measures are of limited interest. But we can create new measures
by perturbing the Gaussian measure $\mu_{C}$ with local interactions.
We cannot study directly the situation where the
interactions are in infinite volume. Instead we put them in a very large volume
which will eventually go to infinity. We have a ratio of two length scales,
one from the size of the diameter of the volume and the other from
the ultraviolet cutoff
in $\mu_{C}$ and this ratio is enormous. The Renormalization group is useful
whenever there are two length scales whose ratio is very large. It permits
us to do a scale by scale analysis and at each step the volume is reduced at
the cost of changing the interactions. The largeness of the ratio is reflected
in the large number of steps to be accomplished this number tending eventually
to infinity. This large number of steps has to be mathematically
controlled.\\
\\{\it Perturbation of the Gaussian measure}\\
\\Let 
$\Lambda_{N}= [-\frac{L^{N}}{2} , \frac{L^{N}}{2}]^{d}\subset {\bf R}^{d}$
be a large cube in $ {\bf R}^{d}$. For any $X\subset \Lambda_{N}$ let
$V_{0}(X, \phi)$ be a local semi bounded function where the 
fields are restricted to the set $X$. Local means that if $X, Y$ are sets with
disjoint interiors then $V_{0}(X\cup Y, \phi)= V_{0}(X)+ V_{0}(Y)$. Consider
the integral ( known as the {\it partition function} in QFT and statistical
mechanics)

\begin{equation} \label{eq-ex.17c}
Z(\Lambda_{N})= \int d\mu_{C}(\phi) z_{0}(\Lambda_{N} ,\phi)
\end{equation}
where
\begin{equation}
z_{0}(X,\phi)= e^{-V_{0}(X,\phi)} \label{eq-ex.17a}
\end{equation}
and

\begin{equation}
d\mu^{(0)}(\Lambda_{N}, \phi)= \frac{1}{Z(\Lambda_{N})}\>  d\mu_{C}(\phi)\> 
e^{-V_{0}(\Lambda_{N},\phi )} 
\end{equation}
is the corresponding probability measure. $V_{0}$ is typically not quadratic in
the fields and therefore leads to a non-gaussian perturbation. For example ,

\begin{equation} 
V_{0}(X,\phi )= \int_{X} dx\> (\xi \> |\nabla\phi(x)|^{2} +    
g_{0}\>  \phi^{4}(x) + \mu_{0}\>  \phi^{2}(x) )
\label{eq-ex.17}
\end{equation}
where we take $ g_{0} >0$. The integral (\ref{eq-ex.17c}) 
is well defined because the sample fields are 
smooth.

We now proceed to the scale by scale analysis mentioned earlier.
Because $\m_{C}$ is an invariant measure of $T_{L}$ we have for the partition
function $Z(\Lambda_{N})$ in the volume $\Lambda_{N}$ 

\begin{equation}
Z(\Lambda_{N}) =\int d\m_{C}(\phi) z_{0}(\Lambda_{N},\phi)
= \int d\m_{C}(\phi)\> T_{L}z_{0}(\Lambda_{N},\phi)  
\end{equation}
The integrand on the right hand side is a new function of fields which 
because of the final scaling live in 
the smaller volume $\Lambda_{N-1}$. This leads to the definition :

\begin{equation} \label{eq-ex.17b}
z_{1}(\Lambda_{N-1},\phi)= T_{L}z_{0}(\Lambda_{N},\phi)                  
\end{equation}
Because $V_{0}$ was local $z_{0}$ has a factorization property for unions
of sets with disjoint interiors. This is no longer the case for $z_{1}$.
Wilson noted that nevertheless the integral is well approximated by an
integrand which does but the approximator has new coupling constants. The
phrase ``well approximated'' is what all the rigorous work is about and
this was not evident in the early Wilsonian era.
The idea is to extract out a local part  and {\it also} consider the 
remainder. The local part leads to a flow of coupling constants and the 
(unexponentiated ) remainder is an irrelevant term. This operation and its
mathematical control is an essential feature of RG analysis.\\
\\Iterating the above transformation we get for all $0\le n\le N$

\begin{equation}
z_{n+1}(\Lambda_{N-n-1},\phi)= T_{L}z_{n}(\Lambda_{N-n},\phi)                  
\end{equation}
After $N$ iterations we get 
\begin{equation}
Z(\Lambda_{N})=\int d\m_{C}(\phi) z_{N}(\Lambda_{0},\phi)
\end{equation}
where $\Lambda_{0}$ is the unit cube. To take the $N\rightarrow \infty$ we have
to control the infinite sequence of iterations. We cannot hope to control
the infinite sequence at the level of the entire partition function.  
Instead one chooses representative coordinates for which the infinite 
sequence has a chance of having a meaning. The coordinates are provided by the 
coupling constants of the 
extracted local part and the irrelevant terms, (an approximate calculation
of the flow of coupling constants is given in Section 3).
The existence of a global trajectory for such coordinates helps us to control 
the limit
for moments of the probability measure ({\it correlation functions}). The 
question of coordinates and the representation of the 
irrelevant terms will be taken up in Section 4. \\ 
\\{\it Ultraviolet cutoff removal }: \\
\\The next issue is ultraviolet cutoff removal in field theory. 
This problem can be put into the earlier
framework as follows.
Let $\epsilon_{N}$ be a sequence of positive
numbers which tend to $0$ as $N\rightarrow \infty$ . 
Following the {\it Remark} after (\ref{eq-ex.2})
we replace the unit cutoff covariance $C$ by the  covariance 
$C_{\epsilon_{N}}$ defined by taking $\epsilon_{N}$ instead of $1$ as 
the lower end point in the integral (\ref{eq-ex.2}). Thus $\epsilon_{N}$
acts as a short distance or ultraviolet cutoff. It is easy to see that

\begin{equation}
C_{\epsilon_{N}}(x-y)= S_{\epsilon_{N}} C(x-y)  
\end{equation}
Consider the partition function $Z_{\epsilon_{N}}(\Lambda)$  in a  cube
$\Lambda = [-\frac{R}{2},\frac{R}{2}]^{d}$ :

\begin{equation} \label{eq-ex.18}
Z_{\epsilon_{N}}(\Lambda)= \int d\mu_{C_{\epsilon_{N}}}(\phi)   \>
e^{-V_{0}(\Lambda,\> \phi,\> \tilde\xi_{N} , \> \tilde g_{N},\> \tilde\mu_{N})}
\end{equation}
where $V_{0}$ is given by (\ref{eq-ex.17}) with $ g_{0}, \mu_{0}$  replaced
by $\tilde g_{N}, \tilde\mu_{N}$. By dimensional analysis we can write

\begin{equation} \label{eq-ex.19}
\tilde\xi_{N} =\epsilon_{N}^{(2[\phi]-d + 2)}\xi, \quad  
\tilde g_{N} = \epsilon_{N}^{(4[\phi]-d)} g , \quad 
\tilde\mu_{N}= \epsilon_{N}^{(2[\phi]-d)} \mu
\end{equation}
where $g, \xi, \mu$ are dimensionless parameters. Now $\phi$ distributed 
according to $C_{\epsilon_{N}}$ equals in distribution $S_{\epsilon_{N}}\phi$
distributed according to $C$. Therefore choosing $\epsilon_{N} = L^{-N}$
we get

\begin{align} \label{eq-ex.20}
Z_{\epsilon_{N}}(\Lambda) & =  \int d\mu_{C}(\phi)   \>
e^{-V_{0}(\Lambda,\> S_{\epsilon_{N}}\phi,\>  \tilde\xi_{N}, \> 
\tilde g_{N},\> \tilde\mu_{N})} \cr
& = \int d\mu_{C}(\phi)e^{-V_{0}(\Lambda_{N},\> \phi, \> \xi \> g ,\> \mu)}
\end{align}
where $\Lambda_{N} = [-L^{N}\frac{R}{2},L^{N} \frac{R}{2}]^{d}$. 
Thus the field theory problem of ultraviolet cutoff removal i.e. 
taking the limit $\epsilon_{N} \rightarrow 0 $ has been reduced
to the study of a statistical mechanical model in a very large volume.
The latter has to be analyzed via RG iterations as before.\\
\\{\it Critical field theories}

As mentionned earlier we have to study the flow of local interactions
as well as that of irrelevant terms. Together they constitute the RG 
trajectory and we have to prove that it exists globally. In general the
trajectory will tend to explode after a large number of iterations
due to growing relevant terms (characterized in terms of the expanding
Wick monomials mentionned earlier). Wilson pointed out that the saving factor
is to exploit fixed points and invariant manifolds of the fixed points by
tuning the initial interaction so that the RG has a global trajectory. This 
leads to the notion of a {\it critical manifold} which can be defined as 
follows. A fixed point will have
contracting and/or marginal attractive directions besides expanding ones.
In the language of dynamical systems the critical manifold is the {\it stable}
or {\it center stable manifold} of the fixed point in question. 
This is determined by a detailed study
of the discrete flow. In the examples above it amounts to fixing the initial 
``mass'' parameter  $\mu_{0} = \mu_{c}(g_{0})$ with a suitable function 
$\mu_{c}$ such that the flow remains bounded in an invariant set. The critical 
manifold is then the graph of a function from the space of contracting and 
marginal variables to the space of $\mu$'s which remains invariant under 
the flow. Restricted to it the flow will now converge 
to a fixed point. All reference to initial coupling 
constants have disappeared. The result is known as a {\it critical theory}.

Critical theories have been rigorously
constructed in a number of cases. Take the standard $\phi^{4}$
in $d$ dimensions. Then $[\phi]=\frac{d-2}{2}$. For $d > 5$ the $\phi^{4}$
interaction is irrelevant and the Gaussian fixed point is attractive with one unstable
direction ( corresponding to $\mu$ ).
In this case one can prove that the interactions
converge exponentially fast to the Gaussian fixed point 
on the critical manifold.  For $d=4$ the interaction is 
marginal and the Gaussian fixed point attractive for $g>0$. 
The critical theory has been constructed by Gawedzki and 
Kupainen in [GK 1985] starting with a sufficiently small coupling constant.
The fixed point is Gaussian ( interactions vanish in the limit)
and the convergence rate is logarithmic. This is thus
a mean field theory with logarithmic corrections as expected on heuristic 
grounds. The mathematical construction of the critical theory in $d=3$ is an 
open problem.
( It is expected to exist with a non-Gaussian fixed point and this is indicated
by the perturbative $\epsilon$ expansion of Wilson and Fisher in 
$ 4-\epsilon $ dimensions ). 
However the critical theory for $d=3$ for $[\phi]= \frac{3-\epsilon}{4}$
for $\epsilon > 0$ held very small has been rigorously constructed by 
Brydges, Mitter and Scoppola in [BMS 2003]. This theory has
a nontrivial hyperbolic fixed point of $O(\epsilon)$. The stable manifold
is constructed in a small neighborhood of the fixed point.
Note that the ( uncutoff) covariance
is Osterwalder-Schrader positive and thus this is a candidate for a non-trivial
EQFT. For $\epsilon =1$ we have the standard situation
in $d=3$ and this remains open as mentioned earlier. A very simplified picture
of the above is furnished by the perturbative computation in Section 3. \\
\\{\it Unstable fixed points}

We may attempt to construct field theories around unstable fixed points.
In this case the initial parameters have to be adjusted as functions of the 
cutoff in such a way as to stabilize the flow in the neighborhood of the fixed
point. This may be called a {\it genuine} renormalization.
A famous example of this is pure Yang-Mills theory in $d=4$ where the Gaussian
fixed point has only marginal unstable directions. Balaban
in a series of papers ending in [B] considered Wilson's lattice cutoff version
of Yang-Mills theory in $d=4$ with initial coupling fixed by the two loop
asymptotic freedom formula. He proved by lattice RG iterations that 
in the weak coupling regime the free energy per unit 
volume is bounded above and below by constants independent of the lattice spacing. 
Instability of the flow is expected to lead to mass generation for observables but
this is a famous open problem. Another example 
is the standard non-linear sigma model for d=2. Here too  
the flow is unstable around the Gaussian fixed point and we can set the 
initial coupling
constant by the 2-loop asymptotic freedom formula. Although much is known
via approximation methods ( as well as by integrable systems methods )
this theory remains to be rigorously constructed as an EQFT .      

Let us now consider a relatively simpler example, that of constructing a 
massive 
super-renormalizable scalar field theory. This has been studied in $d = 3$, 
with $[\phi]=\frac{d-2}{2} = \frac{1}{2}$.
We get $\xi=\tilde\xi,\>  g=L^{-N}\tilde g, \> \mu = L^{-2N}\tilde \mu $ and 
$\tilde g $ is taken to be small. $\xi$ is marginal whereas $g, \> \mu$
are relevant parameters and grow with the iterations. After $N$ iterations
they are brought upto $\tilde g, \> \tilde \mu $ together with remainders.
This realizes the so called massive continuum $\phi^{4}$ theory in $d=3$
and this has been mathematically controlled in the exact RG framework.
This is proved by Brydges, Dimock and Hurd in [BDH 1995] and earlier 
by Benfatto, Cassandro, Gallavotti and others in [BCGNOPS 1980].

\section{The exact RG as a Continuous Semigroup}\label{sec-cont}
\setcounter{equation}{0}

The discrete semigroup defined in (\ref{eq-ex.10} 
of the previous section has a natural continuous counterpart. Just take
$L$ to be a continuous parameter, $L= e^{t}: t\ge 0$ and write
by abuse of notation $T_{t}, S_{t}, \Gamma_{t}$ instead of $T_{e^{t}}$ etc. 
The continuous transformations $T_{t}$
\begin{equation}
T_{t}F = S_{t}\> \mu_{\Gamma_{t}} * F \label{eq-cont.7}
\end{equation}
give a semigroup

\begin{equation}
T_{t}T_{s} =T_{t+s} \label{eq-cont.8}
\end{equation}
of contractions on $L^{2}(d\mu_{C})$ with $\mu_{C}$ as invariant measure.
One can show that $T_{t}$ is strongly continuous and therefore has a  
generator which we will call $\cal L$. This is defined by

\begin{equation}
{\cal L} F =  \lim_{t\rightarrow 0^{+}} \frac{T_{t}-1}{t} F \label{eq-cont.8}
\end{equation}
whenever this limit exists. This restricts $F$ to a suitable subspace
${\cal D(L)} \subset L^{2}(d\mu_{C})$. $\cal D(L)$ contains for example 
polynomials
in fields as well as twice differentiable bounded cylindrical functions. 
The generator $\cal L$ can be easily computed. To state it 
we need some definitions. Define 
$(D^{n}F)(\phi ; f_{1}, \ldots, f_{n} )$ as the $n$-th tangent map at $\phi$
along directions $ f_{1}, \ldots, f_{n}$. The functional
Laplacian $\Delta_{\dot \Gamma}$ is defined by

\begin{equation}\label{eq-cont.9}
\Delta_{\dot \Gamma} F(\phi) = \int d\mu_{\dot \Gamma}(\zeta)\>
(D^{2}F)(\phi ; \zeta, \zeta)
\end{equation}
where ${\dot\Gamma}= u$. Define an infinitesimal dilatation operator

\begin{equation} \label{eq-cont.10}
{\cal D}\phi(x)= x\cdot \nabla \phi(x)
\end{equation}
and a vector field $\cal X$

\begin{equation} \label{eq-cont.11}
{\cal X}F = -[\phi] (DF)(\phi ; \phi)- (DF)(\phi ; {\cal D}\phi )
\end{equation}
Then an easy computaion gives 
\begin{equation} \label{eq-cont.12}
{\cal L} =\frac{1}{2}  \Delta_{\dot \Gamma} + {\cal X}
\end{equation}
$T_{t}$ is a semigroup with $\cal L$ as generator. Therefore 
$T_{t}=e^{t\cal L}$. Let $ F_{t}(\phi)=  T_{t} F(\phi)$. Then 
$F_{t}$ satisfies the linear PDE 

\begin{equation} \label{eq-cont.13}
\frac{\partial F_{t}}{\partial t} = {\cal L} F_{t}
\end{equation}
with initial condition $F_{0}=F$. This evolution equation assumes a more
familiar form if we write $F_{t}= e^{-V_{t}}$, $V_{t}$ being known as the
{\it effective potential}. We get
\begin{equation}  \label{eq-cont.14}
\frac{\partial V_{t}}{\partial t} = {\cal L}V_{t}
 - \frac{1}{2} (V_{t})_{\phi}\cdot (V_{t})_{\phi}
\end{equation}
where

\begin{equation} 
(V_{t}(\phi))_{\phi}\cdot (V_{t}(\phi))_{\phi}=\int d\mu_{\dot \Gamma}(\zeta)
( (DV_{t})(\phi ; \zeta))^{2}
\end{equation}
and $V_{0}=V$. This infinite dimensional non-linear PDE is a version of 
Wilson's {\it flow equation}. 
Note that the linear semigroup $T_{t}$ acting on functions  induces a semigroup
${\cal R}_{t}$ acting non-linearly on  effective potentials giving a trajectory
$V_{t}= {\cal R}_{t} V_{0}$.

Equations like the above 
are notoriously difficult to control rigorously especially
so for large times. However
they may solved in formal perturbation theory when the initial $V_{0}$ is small
via the presence of small parameters. In particular they give rise easily
to perturbative flow equations for coupling constants. They can be obtained to
any order but then there is the remainder. It is hard to control the remainder
from the flow equation for effective potentials in bosonic field theories.
 They require other methods based on the discrete RG . 
Nevertheless these approximate
perturbative flows are very useful for getting a prelimnary view of the flow. 
Moreover their
discrete versions figure as an imput in further non-perturbative analysis.\\ 
\\ {\it Perturbative flow} : \ \ It is instructive
to see this in second order perturbation theory. We will simplify
by working in infinite volume ( no infrared divergences can arise
because $\dot\Gamma (x-y)$ is of fast decrease ). Now suppose that we are
in standard $\phi^{4}$ theory with $[\phi] = \frac{d-2}{2}$
and $d >2$. We want to show that
\begin{equation} \label{eq-cont.14a}
V_{t}=\> \int dx \> \Bigl( \xi_{t} :|\nabla\phi(x)|^{2}: + g_{t}:\phi(x)^{4}: +
\mu_{t}:\phi(x)^{2}: \Bigr) 
\end{equation}
satisfies the flow equation in second order modulo  irrelevant terms provided
the parameters flow correctly. We will ignore field independent terms.
The Wick ordering is with respect to the covariance $C$ of the invariant 
measure. The reader will notice that we have ignored a $\phi^{6}$ term which
is actually relevant in $d=3$ for the above choice of $[\phi]$. This is because
we will only discuss the $d=3$ case for the model discussed at the end of this
section and for this case the $\phi^{6}$ term is irrelevant.
We will assume that $\xi_{t}, \mu_{t}$ are of $O(g^{2})$. Plug in 
the above in the flow equation. $\lambda_{t}^{n,m} :P_{n,m}:$ represent 
one of the terms above with $m$ fields and $n$ derivatives. Because $\cal L$ 
is the generator of the semigroup $T_{t}$ we have

\begin{equation} \label{eq-cont.14b}
(\frac{\partial}{\partial t} -{\cal L})\lambda_{t}^{n,m} :P_{n,m}: =
(\frac{d\lambda_{t}^{n,m}}{dt} - (d-m[\phi]-n)\lambda_{t}^{n,m}) :P_{n,m}: 
\end{equation}
Next turn to the non-linear term in the flow equation and insert the $\phi^{4}$
term ( the others are already of $O(g^{2})$ ). This produces a double integral
of ${\dot\Gamma}(x-y):\phi(x)^{3}::\phi(y)^{3}:$ which after complete Wick
ordering gives
\begin{equation}
-\frac{g_{t}^{2}}{2} 16
\int dx dy\> {\dot\Gamma}(x-y)\bigl(:\phi(x)^{3}\phi(y)^{3}: +
9 C(x-y):\phi(x)^{2}\phi(y)^{2}: + 36 C(x-y):\phi(x)\phi(y):   + 6
C(x-y)^{2}\bigr)
\end{equation}
Consider the non-local $\phi^{4}$ term. We can localize it by writing

\begin{equation}
:\phi(x)^{2}\phi(y)^{2}:= \frac{1}{2} : \Bigl( \phi(x)^{4} +\phi(y)^{4}
-(\phi(x)^{2} - \phi(y)^{2})^{2}\Bigr) :
\end{equation}
The local part gives a $\phi^{4}$ contribution  and the  the last term above 
gives rise to an irrelevant contribution because it
produces additional derivatives. The coefficients are well defined because
$C, \dot\Gamma$ are smooth and $\dot\Gamma(x-y)$ is of fast decrease. 
Now the
non-local $\phi^{2}$ term is similarly localized. It gives a relevant local
$\phi^{2}$ contribution  as well as a marginal $|\nabla \phi|^{2}$ 
contribution.
Finally the same principle applies to the non-local $\phi^{6}$ contribution
and gives rise to further irrelevant terms.
Then it is easy to see by matching that the flow equation is satisfied in 
second order up to irrelevant
terms ( these would have to be compensated by adding additional terms in 
$V_{t}$ )  provided
\begin{align}\label{eq-cont.14c}
\frac{dg_{t}}{dt}& = ( 4-d)g_{t} - a g_{t}^{2} + O(g_{t}^{3}) \cr
\frac{d\mu_{t}}{dt} &  = 2 \mu_{t} - b g_{t}^{2} + O(g_{t}^{3})   \cr
\frac{d\xi_{t}}{dt} & = c  g_{t}^{2} + O(g_{t}^{3})
\end{align}
where $a,b,c$ are positive constants. We see from the above formulae that
up to second order in $g^{2}$ as $t\rightarrow \infty$, 
$g_{t}\rightarrow 0$ for $d \ge 4$. In fact for $d\ge 5$ the decay rate is
$O(e^{-t})$ and for $d=4$ the rate is $O(t^{-1})$.  However to see if 
$V_{t}$  converges we have to also discuss the
$\mu_{t}, \xi_{t}$ flows. It is clear that in general the $\mu_{t}$ flow
will diverge. This is fixed by choosing the initial $\mu_{0}$ to be the
{\it bare critical mass}. This is obtained by integrating upto time $t$ and 
then
expressing $\mu_{0}$ as a function of the entire $g$ trajectory up to time 
$t$. Assume that $\mu_{t}$ is uniformly bounded and take $t\rightarrow \infty$.
This gives the critical mass as

\begin{equation}
\mu_{0}=b\int_{0}^{\infty} ds\> e^{-2s}g_{s}^{2} = \mu_{c}(g_{0})
\end{equation} 
This integral converges for all cases discussed above. With this choice of
$\mu_{0}$ we get

\begin{equation}
\mu_{t}=b\int_{0}^{\infty} ds\> e^{-2s}g_{s+t}^{2}
\end{equation}
and this exists for all $t$ and converges as $t\rightarrow \infty$.
Now consider the perturbative $\xi$ flow. It is easy to see from the above 
that for $d\ge 4$, $\xi_{t}$ converges as $t\rightarrow \infty$. 

We have not discussed the $d=3$ case because the perturbative $g$ fixed point is
of $O(1)$. But suppose we take in the  $d=3$ case
$[\phi]= \frac{3-\epsilon}{4}$  with $\epsilon > 0$
held small as in [BMS 2003]. Then the above perturbative flow equations are easily
modified ( by taking account of (\ref{eq-cont.14b}))
and we get to second order an attractive
fixed point $g_{*}= O(\epsilon)$ of the $g$ flow.
The critical bare mass $\mu_{0}$ can be determined as before and the $\xi_{t}$ flow
converges. The qualitative picture obtained above has a rigorous justification.

\section{Rigorous RG analysis}\label{sec-rigor}
\setcounter{equation}{0}

We will give a brief introduction to rigorous RG analysis in the discrete set 
up of Section 2 concentrating on the
principal problems encountered and how one attempts to solve them. 
The approach we will outline is
borrowed from Brydges, Mitter and Scoppola [BMS 2003]. It is 
a simplification of the 
methods initiated by Brydges and Yau in [BY 1990] and developed further by
Brydges, Dimock and Hurd [BDH 1998]. The reader will find in the selected 
references other approaches to rigorous RG methods such as those of Balaban,
Gawedzki and Kupiainen, Gallavotti and others. We will take as a 
concrete example the scalar field model introduced earlier.\\
\\At the core of the analysis is the choice of good coordinates for the 
partition function density $z$ of Section 2. This is provided by a
polymer representation ( defined below) which parametrizes $z$ by a couple
$(V, K)$ where $V$ is a local potential and $K$ is a set function depending
also on the fields. Then the RG transformation $T_{L}$ maps $(V, K)$ to a new
$(V, K)$. $(V, K)$ remain good coordinates as the volume $\rightarrow \infty$
whereas $z({\rm volume})$ diverges. There exist norms which are suited 
to the fixed point analysis of $(V, K)$ to new $(V, K)$. Now comes the 
important point: $z$ does not uniquely specify the representation $(V, K)$.
Therefore we can take advantage of this non-uniqueness to keep $K$ small in 
norm and let most of the action of $T_{L}$ reside in $V$. This process is
called {\it extraction} in [BMS 2003]. It makes sure that $K$ is an irrelevant
term whereas the local flow of $V$ gives rise to discrete flow equations in
coupling constants. We will not discuss extraction any further. In the 
following we introduce the polymer representation and explain how the RG
transformation acts on it.\\ 
\\To proceed we introduce first a simplification in the set up of Section 2.
Recall that the function $u$ introduced before (\ref{eq-ex.2}) was
smooth, positive definite, and of rapid decrease. We will simplify further
by imposing the stronger property that it is actually of
{\it finite range} : $u(x)= 0 $ for $|x|\ge 1$. We say that
$u$ is of finite range $1$. It is easy to construct such functions.
For example if $g$ is any smooth function of finite range $\frac{1}{2}$ then
$u=g*g$ is a smooth positive definite function of finite range $1$. 
This implies that the
fluctuation covariance $\Gamma_{L}$ of (\ref{eq-ex.4}) has finite range $L$.
As a result $\Gamma_{n}$ in (\ref{eq-ex.7}) has finite range $L^{n+1}$
and the corresponding fluctuation fields $\zeta_{n}(x)$ and $\zeta_{n}(y) $ 
are independent when $|x-y|\ge L^{n+1}$.\\
\\{\it Polymer representation}

Pave ${\bf R}^{d}$ with closed cubes of side length $1$ called 
$1$- {\it blocks} or unit blocks denoted by $\Delta$,  and suppose 
that $\Lambda$ is a large cube consisting of unit blocks. A {\it connected 
polymer} $X\subset \Lambda$ is a closed  connected subset of these unit blocks.
 A polymer
{\it activity} $K(X,\phi)$ is a map $X, \phi \rightarrow \bf R$ where the 
fields
$\phi$ depend only on the points of $X$. We will set $K(X,\phi) =0 $ if $X$
is not connected. A generic form of the partition function density 
$z(\Lambda, \phi)$ after a certain number of RG iterations is

\begin{equation} \label{eq-disc.1}
z(\Lambda)=\sum_{N=0}^{\infty} \frac{1}{N !} \sum_{X_{1},..., X_{N}}
e^{-V(X_{c})}\prod_{j=1}^{N} K(X_{j})
\end{equation}
Here the $X_{j}\subset \Lambda$ are disjoint polymers, $X=\cup X_{j}$, and
$X_{c}=\Lambda \setminus X$. $V$ is a local potential of the form
(\ref{eq-ex.17}) with parameters $\xi, g, \mu$. We have suppressed the 
$\phi$ dependence. Initially the activities $ K_{j}=0$ but they will arise 
under RG 
iterations and the form (\ref{eq-disc.1}) remains stable as we will  see. The 
partition function density is thus parametrised as a couple $( V,\> K )$. \\
\\{\it Norms for polymer activities} : \\
\\Polymer activities $K(X,\phi)$ are endowed with a norm  
$\Vert K(X)\Vert$ which must satisfy two properties:

\begin{align} \label{eq-disc.1a}
{\dot X}\cap {\dot Y} = 0 \Rightarrow 
\Vert K_{1}(X) K_{2}(Y)\Vert & \le \Vert K_{1}(X)\Vert \>
\Vert K_{1}(Y)\Vert \> \cr
\Vert T_{L} K(X)\Vert  \le c^{|X|} \Vert K(X)\Vert
\end{align}
where ${\dot X}$ is the interior of $X$ and $|X|$ is the number of blocks in 
$X$. $c$ is a constant of $O(1)$.
The norm measures ( Fr\'ech\'et ) differentiability properties of the 
activity $K(X,\phi)$
with respect to the field $\phi$ as well as its admissible growth in $\phi$. 
The
growth is admissible if it is $\mu_{C}$ integrable.
The second property above ensures the {\it stability} of the norm under  
RG iteration . For a fixed polymer $X$ the norm is such that it gives rise to 
a Banach space of activities $K(X)$.  
The final norm $\Vert \cdot \Vert_{\cal A}$
incorporates the previous one and washes out the set dependence

\begin{equation} \label{eq-disc.1b}
\Vert K \Vert_{\cal A}= \sup_{\Delta}\sum_{X\supset\Delta} A(X)\>  
\Vert K(X)\Vert
\end{equation}
where $A(X)= L^{(d+2)|X|}$. 
This norm essentially ensures that large polymers have small activities. The 
details of the above norms can be found in [BMS 2003].\\
\\The RG operation map $f$ is a composition of two maps. The RG iteration map 
$z\rightarrow T_{L}z$
induces a map $V\rightarrow {\tilde V}_{L}'$ and a nonlinear map  
${\tilde T}_{L}:\>   K\rightarrow {\tilde K}={\tilde T}_{L}(K)$. 
We then compose this with a ( nonlinear )
{\it extraction} map $\cal E$ which takes out the expanding ( relevant) parts of 
${\tilde K}\rightarrow {\cal E}({\tilde K})=K' $ and compensates the 
local potential $ {\tilde V}_{L}'\rightarrow V'$ such that $T_{L} z$ remains invariant.
We denote by $f$ the composition of these two maps with 
\begin{equation}
V\rightarrow V'=f_{V}(V,K),\quad K\rightarrow K'=f_{K}(V,K)
\end{equation}

\vskip0.3cm

\noindent {\it The map} ${\tilde T}_{L}$ :\\
\\Consider applying the RG map $T_{L}$ to (\ref{eq-disc.1}).
The map consists of a convolution $\mu_{\Gamma_{L}} *$ followed by the
rescaling $S_{L}$. In doing the integration over the fluctuation field $\zeta$ 
we will exploit the independence of $\zeta (x)$ and $\zeta (y)$ when
$|x-y| \ge L$. To do this we pave $\Lambda$ by closed
blocks of side $L$ called $L$-blocks
so that each $L$-block is a union of $1$-blocks. Let ${\bar X}^{L}$  be the 
$L$-closure of a set $X$, namely the smallest union of $L$-blocks containing $X$.
The polymers will be combined into $L$-polymers which are by definition connected
unions of $L$ blocks. The combination is performed in such a way that the new 
polymers are associated to independent functionals of $\zeta$.\\  
\\Let ${\tilde V}(X,\phi))$, to be chosen later,
be a local potential independent of $\zeta$. For the coupling constant 
sufficiently small there is a bound

\begin{equation} \label{eq-disc.2a}
\Vert e^{-V(Y)}\Vert \le 2^{|Y|}
\end{equation}
We assume that $\tilde V$ is so chosen that  the same bound holds when $V$ is 
replaced by $\tilde V$. Define

\begin{equation} \label{eq-disc.2}
P(\Delta, \zeta, \phi)= e^{-V(\Delta, \zeta +\phi)} - e^{-{\tilde V} (\Delta, \phi)}
\end{equation}  
Then we have 

\begin{equation}
e^{-V(X_{c} , \zeta +\phi)}  = e^{-V({\bar X}_{c} , \zeta +\phi)} = 
\prod_{\Delta\subset {\bar X}_{c}} (e^{-{\tilde V} (\Delta, \phi)}     +
P(\Delta, \zeta, \phi))
\end{equation}
where ${\bar X}_{c}$ is the closure of $X_{c}$.
Expand out the product and insert  into the representation (\ref{eq-disc.1})
for $z(\Lambda, \zeta +\phi))$. We then rewrite the resulting sum in terms
of $L$-polymers. The sum splits into a sum over connected components. Define
for every connected  $L$-polymer $Y$

\begin{equation} \label{eq-disc.4}
{\cal B}K(Y)= \sum_{N+M\ge 1} \frac{1}{N ! M !}
\sum_{(X_{j}), (\Delta_{i})\rightarrow Y} e^{-{\tilde V}(X_{0})}
\prod_{j=1}^{N} K(X_{j})\prod_{i=1}^{M} P(\Delta_{i} )
\end{equation}
where $X_{0}= Y\setminus (\cup X_{j})\cup (\cup \Delta_{j})$ and 
the sum over the distinct $\Delta_{i}$, and disjoint $1$-polymers $X_{j}$ is
such that their $L$-closure is $Y$. (\ref{eq-disc.1}) now becomes

\begin{equation}
z(\Lambda )=\sum \frac{1}{N !}\sum_{Y_{1},..,Y_{N}} e^{-{\tilde V}(Y_{c})}
\prod_{j=1}^{N} {\cal B}K(Y_{j})
\end{equation}
where the sum is over disjoint, connected closed $L$-polymers. We now perform
the fluctuation integration over $\zeta$ followed by the rescaling. Now 
${\tilde V}(Y_{c})$ is independent of $\zeta$. The $\zeta$ integration sails
through and then factorizes because the $Y_{j}$ being disjoint closed 
$L$-polymers
they are separated from each other by a distance $\ge L$. The rescaling brings
us back to $1$-polymers and reduces the volume from $\Lambda$ to 
$L^{-1}\Lambda$. Therefore

\begin{equation} \label{eq-disc.5}
z'(L^{-1}\Lambda)= T_{L}z(\Lambda )=
\sum \frac{1}{N !}\sum_{X_{1},..,X_{N}} e^{-{\tilde V}_{L}(X_{c})}
\prod_{j=1}^{N} (T_{L}{\cal B}K)(X_{j})
\end{equation}
where the sum is over disjoint $1$-polymers, $X_{c}= L^{-1}\Lambda\setminus X$.
By definition ${\tilde V}_{L}(\Delta) = S_{L}{\tilde V}(L\Delta)$ and
$(T_{L}{\cal B}K)(Z)= S_{L}\mu_{\Gamma_{L}}*{\cal B}K(LZ)$.
This shows that the representation (\ref{eq-disc.1}) is stable under iteration 
and furthermore gives us the map 

\begin{align} \label{eq-disc.5a}
V \rightarrow {\tilde V}_{L} \cr
K \rightarrow {\tilde K}=  {\tilde T}_{L}(K)= T_{L}{\cal B}K
\end{align}
The norm boundedness of $K$ implies that ${\tilde T}_{L}(K)$ is norm bounded.
We see from the above that a variation in the choice of $\tilde V$ is 
reflected in the corresponding variation of $\tilde K$. The extraction map
$\cal E$ now takes out from $\tilde K$ the expanding parts
and then compensates it by a change of ${\tilde V}_{L}$ in such a way that
the representation (\ref{eq-disc.5}) is left invariant by the simultaneous 
replacement ${\tilde V}_{L}\rightarrow  V',\> 
{\tilde K} \rightarrow K'= {\cal E}( \tilde K) $. The extraction map
is non-linear. Its linearization is a subtraction operation and this dominates
in norm the non-linearities, [BDH 1998].\\
\\ The map $V\rightarrow {\tilde V}_{L}\rightarrow V'$ leads to a discrete
flow of the coupling constants in $V$. It is convenient to write 
$K= K^{pert} +R$ where $R$ is the remainder. Then the coupling constant flow
is a discrete version of the continuous
flows encountered in Section 3 together with remainders which are controlled
by the size of $R$. In addition we have the flow of $K$. The discrete flow of
the pair $({\rm coupling\> constants},\> K)$ can be studied in a Banach space 
norm. Once one 
proves that the non-linear parts satisfy a Lipshitz property the discrete flow
can be analyzed by the methods of stable manifold theory of dynamical systems
in a Banach space context. The reader is referred to the article [BMS 2003]
for details of the extraction map and the application of stable manifold 
theory in the construction of a global RG trajectory.

\section{Further topics}\label{sec-other}
\setcounter{equation}{0}

1. {\it  Lattice RG methods}: Statistical mechanical systems are often 
defined on a lattice. Moreover the lattice provides an ultraviolet cutoff for 
euclidean field theory compatible with Osterwalder-Schrader positivity. The 
standard lattice RG is based on Kadanoff-Wilson block spins. Its mathematical 
theory and applications have been developed by Balaban starting with [B 1982] 
and Gawedski and Kupiainen starting with [GK 1980]. This leads to multiscale 
decompositions of the Gaussian lattice field as a sum of independent 
fluctuation fields on increasing length scales. Brydges, Guadagni and Mitter 
[BGM 2004] have shown that standard Gaussian lattice 
fields have multiscale decompositions as a sum of independent fluctuation 
fields with the finite range property introduced in section 4. This permits 
the development of rigorous lattice RG theory in the spirit of the continuum 
framework of the previous section.\\
\\2. {\it Fermionic field theories}: Field theories of interacting fermions
are often simpler to handle than bosonic field theories. Because of statistics
fermion fields are bounded and perturbation series converges in finite volume
in the presence of an ultraviolet cutoff. The notion of studying the RG 
flow at the level of effective potentials makes sense. At any given scale
there is always an ultraviolet cutoff and the fluctuation covariance being
of fast decrease provides an infrared cutoff. This is illustrated by the
work of Gawedzki and Kupiainen [GK 1985] who gave a non-perturbative 
construction in the weak effective coupling regime of the 
RG trajectory for the Gross-Neveu model in two dimensions. This is an 
example of a model with an unstable Gaussian fixed point where the
initial coupling 
has to be adjusted as a function of the ultraviolet cutoff consistent with 
ultraviolet asymptotic freedom.

\section{Acknowledgements} 

I am grateful to David Brydges for very helpful comments on a prelimnary
version of this article.

%\bibliography{references}

\end{document}